Making Physics Courses Accessible for Blind Students: strategies for course administration, class meetings and course materials

Megan Holt[1], Daniel Gillen[1], Chelsea Cook[3], Christa Hixson Miller[2], Sacha D. Nandlall[1], Kevin Setter[1], Cary Supalo[4], Paul Thorman[1], Suzanne Amador Kane[1]*

[1]Physics Department, Haverford College, Haverford PA 19041; [2] Technology-enhanced Learning and On-line Strategies, Virginia Tech, Blacksburg, VA 24061; [3] Physics Department, Virginia Tech, Blacksburg, VA 24061; [4] Educational Testing Service, Princeton NJ 08540 & Independence Science, West Lafayette, IN 47906.

* Corresponding author

## Introduction

The Americans with Disabilities Act (ADA) mandates that U.S. institutions of higher education provide "reasonable accommodations" to students with disabilities to ensure equal educational opportunities. However, despite the key role of physics as a gateway to Science, Technology, Engineering and Mathematics (STEM) studies, only limited resources exist for teaching physics to students who are blind or visually impaired. Here we share lessons from our experience creating an accessible physics curriculum for a blind physics major. The authors include the student himself, a blind physics B.S. who graduated from a different institution, a PhD chemist and consultant on STEM accessibility who is himself blind, and several sighted educators and course assistants who worked regularly with the students. This article focuses on issues for which instructors are responsible: how to make class meetings, curricular materials, tutorials and demonstrations accessible (as opposed to accommodations determined at an administrative level, such as additional time on tests). An online appendix provides additional resources and specifics to guide actual implementation of these ideas, including a guide to further reading.

Once an institution learns that a blind student will enroll in a physics course, the course instructor and the institutional disabilities coordinator should meet to discuss course logistics well before the semester begins (ideally, over a month or two in advance to allow sufficient lead time). They should begin the process of creating an effective instructional and support team, ensuring key assistive technologies are in place, making all class meetings accessible, and preparing accessible course materials [1].

The most fundamental decision is whether to use individualized instruction, in which the student and instructor meet in separate one-on-one tutorials, or mainstreaming, in which the student attends regular class meetings with other students. We primarily used mainstreaming, supplemented with one-on-one instruction. Instructors should work with the student in question to determine which approach is most suitable given their students' individual needs and the available institutional resources.

**Assembling the instructional team**

A blind student who participates in mainstreamed class meetings should have access to one or more persons who act as an in-class assistant and a tutor outside of class. The in-class assistant ensures all class materials are accessible to the student in real-time while also playing the traditional role of a course tutor when appropriate. For example, they might clarify mathematical notation in a complex equation, describe figures drawn on a chalkboard, or explain visual elements in an interactive demonstration. The tutor provides accessibility help as well as playing the traditional role of a course tutor when appropriate. Neither of these roles can be filled by students currently taking the course, since both positions require advanced familiarity with the course material. We had success hiring either students who had already studied the course material, or part-time instructors with a degree in physics or a related field.

Because locating suitable tutors and other class assistants often requires a specialized search to find the right combination of technical expertise and communications skills, the instructor should share the necessary job description well in advance with the disabilities coordinator. This should include the background knowledge (physics, mathematics and other skills) required for each course. Contacts for locating qualified candidates include local graduate and science education programs, disability coordinators at nearby institutions, and private tutoring websites; the student may also know about tutors or organizations that proved effective in the past. More details and a sample job advertisement are provided in the online appendix.

As soon as possible, all individuals involved in course instruction should receive training in best practices for creating accessible lectures and course materials (see below for details). Throughout the semester, we held regular meetings attended by the entire instructional team to discuss the student's progress, exam content, instructional methods and accessibility issues. Weekly check-ins allowed the lead course instructors to keep the disabilities coordinator informed of the student's progress and any new issues.

This process places a premium on continuity of the instructor and teaching assistants who work with the student who is blind. Maintaining that continuity may be challenging when the student transitions between courses, so it is crucial that the instructional team keep careful records throughout each course to help train new instructors, tutors, and course assistants.

Financing support positions is institution and location specific.  In the US, financial support for college students with disabilities depends on first the state they are from and then the disability. For example, the Virginia Department for the Visually Impaired handles support requests on a case by case basis. Such support may include, orientation and mobility training on the college campus, sponsoring remedial course work that is required by the college, assistive technology training, textbook reimbursement, etc. (https://www.vdbvi.org/).

**Making class meeting places accessible**

The rooms used for class meetings need to be chosen with several considerations in mind.  Poor acoustics or background noise from a hall or ventilation system can make concentration difficult for a student who relies primarily on hearing.  Seating arrangements should allow the student and class assistant to talk in a low voice without unduly disrupting the class.  To facilitate use of assistive technologies, the student should be provided with a table-and-chair setup instead of a tablet arm chair.  Finally, because the student needs time to plan a route to any class meetings, any location changes need to be conveyed as far in advance as possible.

**Preparing accessible course materials**

All course materials that are inherently visual need to be converted into an accessible representation: either audible (speech- or sound-based) or tactile (touch-based) formats, depending on which works best for the student and material in question. A variety of printers and other devices can create "tactile graphics" using patterns of raised dots and other features to represent graphical information and figures from textbooks and other curricular materials. Several programs and devices exist for scanning and converting printed text into either synthetic speech or tactile output. Some blind students are also literate in Braille, a tactile writing system for representing print-based systems of literary, technical and other writing. Braille represents letters and other printed symbols using rectangular arrays of six or eight raised dots. The reader detects Braille symbols by moving the fingertips from left to right across a page or line.

Mathematical and scientific symbols are represented in Braille notations that depend on country. In the United States, the Braille Authority of North America (BANA) is the de facto authority on which Braille codes are used. In 2012, BANA adopted the Unified English Braille code (UEB), which is also used in the UK and other countries. Although UEB includes instructions for the transcription of technical material, BANA has resolved to continue using "Nemeth Code Braille for Mathematics and Science Notation", which we therefore reference in this paper.

As early as possible, the disabilities coordinator, instructor and student should determine the tools to be used in preparing course materials with the goal of delivering them to the blind student at the same time as the rest of the class. While the disabilities coordinator can help identify and provide many in-class accessibility resources, we found that the specific needs and technical nature of physics courses meant that instructional staff often needed to help with this process.  We now describe the two main ways for blind individuals to access documents.

**Screen-reading software:**  For blind students who are not adept Braille readers, some course materials may be better suited for conversion into an audible format.  In addition to using traditional audiobooks or having an assistant read course materials aloud, there are several technologies available for converting text to speech and Braille on demand.  JAWS (Job Access With Speech) screen-reading software (Freedom Scientific, St. Petersburg, Florida) can render text generated by many computer programs on Windows devices (Microsoft Corp., Redmond, WA) into synthetic speech and Braille formats. The same company also produces and distributes MAGic, a screen magnification program suitable for users with low vision. For Apple devices, the current iOS and MacOS (Apple Corp. Cupertino, CA) have an integrated screen reader called VoiceOver, which converts text to either speech or Braille in a way analogous to JAWS for Windows. Some of the Braille devices discussed in the next section also feature built-in synthetic speech that can produce an audible version of the text.  Finally, smartphone applications are being developed for small-scale optical character recognition (OCR) and color identification to access certain printed or visual information using the built-in camera.

When encountering an image, a screen reader will read aloud any alternative text (alt tag) that is available. In HTLM5 based content, such as a learning management system, the alt tag is provided by the content creator who may be the instructor. In document formations, such as Microsoft Word, PowerPoint and PDF, the alternative text is again created by the content creator or instructor. (For more information, consult these web resources:

http://diagramcenter.org/making-images-accessible.html, https://webaim.org/techniques/word/, https://webaim.org/techniques/powerpoint/, and https://webaim.org/techniques/acrobat/).

A screen reader has some ability to read math aloud depending on the operating system, the screen reader, the application being read, and the file format of the content. For example the equation for distance with respect to time, velocity and acceleration would be read as, "d equals v subscript I t plus one-half a t  squared") [2]. Benetech's Math Support Finder can help teachers and users determine requirements to ensure proper speech output:  http://msf.mathmlcloud.org/.

**Preparing tactile course materials:**  Because library reserve readings and course-related websites are often inaccessible, blind students must rely on textbooks for a greater share of their studies than do their sighted classmates.  Unfortunately, even commonly-used textbooks and curricular materials often are not available in Braille, and therefore need to be prepared specially. The original materials even may need to be transcribed manually; this is especially likely for work featuring mathematical notation and complex diagrams.  Thus, producing accessible textbooks and course materials using Braille and tactile graphics can take weeks or even months. For this reason, institutions should consider allowing students who require accessible documents to preregister early so the inventory of course materials can be compiled as soon as possible. The course instructor should provide the disabilities coordinator with a complete and comprehensive list of required materials as soon as possible, ideally well before classes begin. This list should include all printed, graphical, and web-based course materials such as textbook sections, handouts, worksheets, homework assignments, and laboratory manuals.   Converting materials into Braille and tactile graphics can be done either using external specialty services or in-house resources, depending on the nature and context of the material. Commercial conversion services are a good choice for lengthier materials (e.g., textbooks) when a trained Braille Nemeth transcriber is unavailable, as such materials are time consuming to covert in-house. In-house conversion is better suited to small-scale, highly customized, or time-sensitive resources (e.g., homework and exams.)

Students can access Braille and tactile graphics in a variety of ways. A hardcopy printout in book form can provide a permanent record that integrates Braille and graphics; however, these documents can be cumbersome and easily damaged. Many electronic devices that function as native Braille computers or notetakers are available. Their refreshable Braille display can output Braille text--typically one line at a time--by means of arrays of piezoelectric crystals with metal or plastic pins that raise to represent the dots of Braille characters. These devices also have a Braille keyboard, which is useful for generating notes and other written coursework. Our student co-author used the Braillenote Apex BT 32 notetaker (Humanware, Drummondville, QC, Canada).

A variety of printers exist that output Braille and other raised renderings for producing tactile printouts. These are made either by embossing dots or by heating special paper to create swell-form graphics on a pixel-by-pixel basis. Some embossers only print Braille text, while others can also produce tactile graphics (diagrams, plots, etc.). Some also print conventional ink on tactile diagrams for the benefit of sighted users. Ideally, there should be two embossing printers: one available to everyone involved in course preparation and grading, and a second for the student's independent, personal use. The appendix provides additional details about the various tactile output options for Braille and graphics, along with specific advice on preparing documents for conversion.

**Online learning management and homework systems**

The institution's online Learning Management System (LMS) (e.g., Blackboard, Moodle, Canvas, etc.) also may need to be adapted to enable the exchange of accessible curricular materials. The ability of the LMS to interface with assistive technologies such as screen readers are particularly important. The disabilities coordinator should work with the information technology staff to resolve such technical issues before the semester begins. The instructor should also be sure to provide accessible versions of any online course assessment and evaluation tools that are not accessible to blind students via the assistive technologies described above [3].

Many introductory physics courses also utilize online homework systems such as MasteringPhysics (Pearson Higher Education, Upper Saddle River, NJ) and WebAssign (WebAssign/Cengage, Rayleigh, NC). While these systems may be partially accessible via screen reader, the physics instructor should be aware of the considerable time and effort required for a blind student to independently input solutions and formulae into online homework software. Some of our instructors found that using separate files containing accessibly formatted homework problems--such as plain text or Microsoft Word with embedded MathType (Design Science, Long Beach, CA)--required less effort for the student, grader, and instructor than online homework systems.

**Assignments and exams**

Exams, problem sets, and other assignments can be produced using an embossing printer to print tactile text, equations and graphics. Where problems had not already been converted into Braille in the textbook for some homework assignments, we created an accessible electronic version with the assistance of the disabilities coordinator's staff. The student could then read exam materials as tactile outputs, and create solutions using a Braille entry device (e.g., a notetaker) and tactile drawing tools. These were converted back into a visual format for grading and review by the instructors and teaching assistants, a task facilitated by using an embossing printer that can ink-print accompanying text.

As an alternative, blind students may complete assignments (homework, tutorials, etc.) and exams orally, with an instructor or teaching assistant reading out the problems and the student responding with oral answers and, in some cases, tactile sketches or other aids.

**Making lectures, discussion sections and other meetings accessible**

To ensure that blind students gain the same value from classroom lectures as sighted individuals, two issues are paramount: instructor training in best practices for communicating with blind students and provisions for effective note-taking by the student.

The instructor can make small accommodations in how material is presented to make class more accessible to blind students without detracting from a preferred personal style. The key point is that everything written on the board also needs to be conveyed orally. Instructors usually leave considerable board content unspoken and implicit during lecture, such as: which symbols in an equation are or are not contained within parentheses; the beginnings and ends of sub-expressions (numerators and denominators of fractions, exponents, etc.); symbols distinguished by color; key points enclosed with boxes; arrows used to indicate the flow of logic in an argument; and directional terms (this, that, here, there, etc.) accompanied by gestures [4-5].

It therefore helps to provide the student with a copy of key equations and diagrams used in class ahead of time. For best quality, tactile printouts can be prepared in advance of lecture, most easily using existing figures adapted from the tactile print course materials. Tactile drawing boards allow the teaching assistant to create real-time raised-line drawings using a pen-like stylus, resilient pad, and special drawing media (e.g., paper or plastic film). A blind student can feel the resulting pattern with the fingertips, similar to how one might read Braille. Because these devices have inherently low resolution, tactile drawing boards work best for producing quick, approximate sketches. See the online appendix for more details about various options for making real-time tactile drawings.

Accessible versions of lecture demonstrations and other tactile aids are important when information is conveyed visually in 3D. A recent article describes KitFis, a classroom kit that facilitates the construction of tactile representations of common illustrations used in introductory physics courses [6]. 3D tactile aids made of pipe cleaners, straws, sticks or cardboard also can help explain key spatial relationships such as the geometry of the vector cross product. Real-time clarification is often necessary: an extreme point on an object's surface can be interpreted as a maximum or minimum, depending on orientation. Sets of simple mechanical objects such as masses and springs mounted on a board, an object with a nontrivial center of mass, as well as simple levers and balances, can be used to illustrate concepts presented in lecture [7].

Taking Braille notes in class can help promote class engagement, synthesis and listening skills. However, The fast pace and technical content of a typical physics class make it challenging for a blind student to take thorough notes while also paying attention and comprehending the material, asking a teaching assistant questions, or following descriptions of graphical materials. Therefore, we recommend always providing the student with class notes prepared by the instructor or a sighted assistant that include all materials written on the board. One effective strategy is to have these notes recorded by a teaching assistant using Microsoft Word, with equations rendered in MathType. This file can be emailed to the student shortly after the end of class, for conversion into electronic Braille. Some students also find audio recordings of lectures helpful. While systems exist for producing and streaming audio or audiovisual recordings, a general-purpose voice recorder suffices for most purposes when accompanied by complete notes. In our case, the student captured audio recordings of the lectures using a VictorReader Stream (Humanware), which is designed specifically for blind and visually-impaired individuals.

**One-on-one tutoring**

It can be especially challenging for blind students to seek help from other students, locate a tutor or utilize campus academic resources. We found that having a tutor delegated to help in completing assignments and studying for exams helped address this gap, resulting in improved learning outcomes and work quality. Ideally, this tutoring should be conducted in a separate, quiet room to minimize distractions. The student performed most work on a Braille notetaker device connected to a computer monitor, providing the tutor with a real-time video display of all text.

When a tutor could not be physically present, remote tutoring sessions could be conducted using a laptop equipped with Skype (Skype Communications, Luxembourg City, Luxembourg), the JAWS screen reader, and a video capture device (Epiphan Video, Palo Alto, CA) that can convert the video feed from the Braille notetaker into the equivalent of a Skype webcam feed. This allowed the tutor to talk with the student while simultaneously viewing the Braille content.

**Accessible scientific computing**

Currently, blind students face challenges in finding accessible tools for scientific computing. One of the most accessible options is the data collection and analysis software LoggerPro (Vernier, Beaverton, OR), which is compatible with JAWS with an appropriate JAWS script file available from Independence Science (West Lafayette, IN). This compatibility allows a blind student to use a set of JAWS shortcut keystrokes to collect and analyze data during experiments. In Microsoft Excel, basic spreadsheet functionality is relatively accessible using a screen reader, although more advanced dialogs and functions (e.g. performing regressions) can be challenging. MATLAB R2018a (Mathworks, Natick, MA) is accessible when run using a screen reader in command-line only mode (using the *-nodesktop* option); however, the lack of accessibility of the code editor meant that code had to be written in a separate text editor. The command-line debug commands (*dbstop*, *dbclear*, and *dbstatus*) were somewhat helpful in this regard, but debugging errors remained significantly more challenging than for a sighted user. At the time of writing, Mathworks is working on a plan of action to improve the accessibility of its products using

MATLAB Online as a platform to improve keyboard accessibility and on-call support. Most of the dialogs and navigation in Wolfram Alpha and Mathematica 11.2 (Wolfram Corporation, Champaign, IL) as well as the plotting and data analysis program Origin (Origin Lab, Northampton, MA) were not easily accessible to blind user without the guidance of a course assistant. To address this issue for Wolfram software, Kyle Keane, PhD (Education and Accessibility Consultant, Wolfram Research Inc.) recommends that blind users learn to use the Wolfram Language interpreter through the text-based wolframscript interface. Using this interface, users can perform every core computation that can be done in Mathematica. This interface enables users to produce results from their computations in many formats, including ASCII math, plain text, HTML, SVG, MathML, LaTeX, and many other formats. (See the online appendix for resources.)

**Classroom demonstrations**

We also identified accessible classroom demonstrations and laboratory exercises (a topic treated in more depth in the electronic appendices).  For example, tactile electronic circuits can be constructed using Snap Circuits (Elenco Electronics, Wheeling, IL) components labeled in Braille. Some experiments in oscillations and waves are inherently tactile and/or audible. For example, the student could easily locate the resonance of vibrating strings or Chladni plates by listening and feeling the vibrations while varying the driving frequency.  Indeed, many physical phenomena cannot be directly sensed visually.  For example, all students must use detectors to measure the inverse-square law, absorption of radiation by materials, atomic emission spectra, etc.

In "sonification", 2D graphical data (such as x-y plots) can be transformed into audible representations. This might involve converting the data into a sound file in which the independent variable becomes time, and the dependent variable is represented by different frequencies or intensities of sound. For example, one might sonify the exponential decay of the activity of a short-lived radioactive sample. Programs with built-in functions for sonifying data include LoggerPro (Vernier Software & Technology, Beaverton, OR), MATLAB (via the sound or soundsc functions) [8] and Mathematica (via the EmitSound function). For demonstrations or labs, LoggerPro software can be used to sonify slowly-changing detector signals or recorded data for position, velocity, voltage, etc. using the "Continuous Tone Meter. For example, the sonified signals of motions of coupled harmonic oscillators on an air track gave a "pendulum" mode signal that was easily distinguished from that due to "breathing" mode oscillations. Sonification also provides a useful way to represent astrophysical data [9-10].

**Scientific calculators**

Enhanced versions of Texas Instruments scientific calculators used in many high school and college courses allow students to interact via audible cues and sonified data, including the Orion TI-30XS MultiView Talking Scientific Calculator and TI-84 Plus Talking Graphing Calculator (American Printing House for the Blind). These devices have standard display, keypad and calculator functionality. Synthetic speech can be used to access all device features, including symbolic input and output. Students also can use these devices quietly via keypad input, haptic (vibrational) feedback, and audio output through earphones. For the TI-84 Plus, SonoGraph software (Orbit Research, Wilmington, DE), makes graphs accessible by sonification and haptic output. For example, shifting from left to right along the graph slides the output between the left and right stereo headphone channels, and audible cues indicate important features such as axis crossings and which quadrant of the x-y plane is being explored.

## Conclusion

To make physics accessible to blind students, instructors should allow considerable lead time for all activities, assemble and train the necessary instructional team, prepare accessible course materials, use key assistive technologies and best practices for conveying content in class, and ensure all computer tools used are as accessible as possible. We found that *all* students benefit when the instructor clarifies concepts in class or use auditory and tactile aids, thus demonstrating one of the tenents of Universal Design for Learning (UDL) [11]. Sighted students emphasized enthusiastically how much they appreciated the additional tactile and 3D graphical aids described herein, as well as the additional care instructors devoted to oral explanations of mathematics and graphics. This effort stimulated us to take a fresh look at novel ways to communicate physics to all audiences.

## Acknowledgements

We would like to thank Anthony Bafile, Sherrie Borowsky, Carr Everbach, Steve Fabiani, Johnny Gonzalez, David Lippel, Gabriela Moats, Sharon O'Neill, John Simonetti, Richard Webb and Beth Willman for their help and insights throughout this effort.

**Online appendix: Resources for teaching physics to blind students to be submitted to Electronic Physics Auxiliary Publication Service (EPAPS)**

**A1. Organizations, conferences and journals**

- National Federation for the Blind has a website https://nfb.org with many resources, including assistive technology products (e.g., a talking measuring tape, Braille meter sticks and rulers, Braille labeling products, a liquid tactile marker.)

- The American Printing House for the Blind (http://www.aph.org/) provides information concerning a federal quota system that can fund the purchase of some instructional materials through the Act to Promote the Education of the Blind, as well as resource for assistive technologies, including lists of tactile printers and companies that perform Braille transcription

- Perkins School for the Blind offers a variety of resources including their online Assistive Technology Store http://www.perkinsproducts.org/store/en/

- California State University, Northridge (CSUN) Assistive Technology Conference meets regularly to provide a venue for learning about the latest developments in this field: http://www.csun.edu/cod/conference (accessed 9/25/2017).

- The U.S. Federal Quota program (the Act to Promote the Education of the Blind) provides financial support for purchasing accessibility aids:  http://www.aph.org/federal-quota/

- CSUN Center on Disabilities, *Journal on Technology and Persons with Disabilities* (accessed 9/25/2017) http://scholarworks.csun.edu/handle/10211.3/125007 .

- *The Physics Teacher* has published several papers on this topic that can be found using the built-in search engine.  Many of these are included in the extended bibliography below.

## A2. Extended Bibliography

Teach Access website with resources (grants, teaching materials, best practices lists, tutorials for accessible design of websites and mobile apps, etc.) for STEM education (http://teachaccess.org/, accessed 4/11/2018).

"Teaching and learning physics with 3D printing"

(https://www.thingiverse.com/groups/teaching-with-3d-printing/topic:2399, accessed 9/25/2017)

DenneDesigns (3D prints for teaching mathematics)

(https://www.thingiverse.com/dennedesigns/about, accessed 4/11/2018).

Mike Tomac, Cricket Bidleman, and Dan Brown, "Homemade wooden vernier scales for use by blind students," *Phys. Teach*. 54(5), 285-287 (2016).

Nicholas P. Truncale, and Michelle T. Graham, "Visualizing sound with an electro-optical eardrum," *Phys. Teach*. 52, 76-79 (2014).

U.S. Department of Education, National Center for Education Statistics (Chapter 3, *Digest of Education Statistics*, 2013, 2015-011, 2015).

## A3. Hiring teaching assistants and tutors

While it would certainly be advantageous for a candidate to possess experience in teaching students with visual impairments, we have found that other attributes such as communication skills, depth of knowledge, and willingness to adapt to new teaching methods are more important. For this reason, if the student's only disability is vision-related, we would recommend omitting the disability from the title of a job posting (e.g. "Physics tutor for a college student" instead of "Physics tutor for a blind college student"), so as not to deter qualified candidates who might otherwise believe that experience working with disabled students is required for the position. Similarly, when accessibility-related issues are mentioned in the position description, they should be accompanied by a note specifying that the person can receive all of the necessary training, and that such experience is desirable but not required.

**A3.1 Sample job posting for a course assistant or tutor**

This is a sample job description for recruiting a course assistant or tutor for accessibility in a physics class, based on the one we used at our institution.

*Physics Tutor for a College Student*

ABC College is looking for someone with a background in physics and calculus to act as a tutor for an undergraduate student during the Fall 2017 semester. The student is blind and needs assistance in the physics course "PHYS 123: Introductory Mechanics".

Experience teaching students with vision impairments or other disabilities is desirable but not required. A background in physics and teaching/tutoring are of primary importance. The ideal candidate would have a Master's degree in physics or a closely related field. The position requires the ability to come to ABC's campus in the town of XYZ, which is accessible via public transit.

Responsibilities are as follows for a total time of about 15-20 hours per week:

Attend the PHYS 123 course with the student on Mondays, Wednesdays and Fridays 9:30-10:30 AM. Take notes and explain any visual components to the student.

Tutor the student in course material and review lecture notes as well as mathematical concepts/problems related to the course.

Convert electronic documents and create tactile graphics for each course. You will receive training on how to use the relevant assistive technologies. Some of this work can be done remotely from home.

The position will begin on September 6 and last through December 16. Compensation is generous and commensurate with experience. Please contact J. Doe (j.doe@abc.edu) with a resume as soon as possible to discuss specifics.

## A4. Preparing accessible course materials

The next two sections focus on two types of accessibility conversion that arise frequently in physics classes: rendering accessible mathematical symbols, equations, and graphics, as well as the recommended workflow for creating accessible documents containing mathematical notation.

**A4.1 Using Braille to make mathematical notation accessible:** The *Nemeth Braille Code for Mathematics and Science Notation* (hereafter Nemeth Code), is the official Braille code used for technical notation in English. With the exception of a small number of spatial formatting conventions defined by the Nemeth Code, all Braille codes are inherently serial, linear formats, in which even two-dimensional information such as matrices and fractions are encoded as linear sequences of symbols. For a discussion of the pedagogical complications introduced by this constraint, we refer readers to Parry et al. (1997).

**A4.2 Workflow for document creation**

Most non-technical materials containing only text in common ASCII formats (e.g. plain text or Microsoft Word files) are automatically accessible using the technologies mentioned previously (JAWS screen reader on Windows computers, VoiceOver on Apple devices, and Braille notetakers). However, physics course materials often use special notation such as mathematical symbols and Greek letters. The two most popular formats for creating these materials are Microsoft Word files with Equation Editor objects and non-ASCII characters, as well as PDFs generated from LaTeX (latex-project.org) files. Unfortunately, both formats require further conversion to be accessible. Microsoft Word Equation Editor objects are not accessible on devices such as Braille notetakers, and non-ASCII characters also sometimes do not render properly on these units; the same compatibility issues also hold for LaTeX generated PDF files. Conversely, Microsoft Word with MathType works well both with transcription software (e.g., Duxbury and Tiger) and with screen reading software. Direct conversion from LaTeX to Braille is complicated further by the fact that LaTeX is infinitely customizable. Any assistive technology, refreshable braille display, speech output, or transcription software will have to be adapted to account for any custom elements created by the author. By contrast, proprietary software like Microsoft Word is more predictable, making it more likely that its output will function with refreshable braille display, speech output, or transcription software. Moreover, even though LaTeX source code is plain text and is therefore accessible in the same way that any other plain text document would be, reading and understanding complex equations from the LaTeX source can be significantly more difficult compared to an equation expressed in Nemeth Code or the typeset version that a sighted student would see. More information and updated advice can be found on the Blindmath listserver at

http://www.nfbnet.org/mailman/listinfo/blindmath_nfbnet.org .

To address these shortcomings, we developed a workflow for translating technical documents that makes use of several additional software applications. First, we used MathType software (Design Science, Long Beach, CA), which is designed to combine powerful equation typesetting capabilities with a graphical interface (the native Microsoft Word Equation Editor was originally adapted from an early version of MathType). MathType includes an add-on for Microsoft Word, with the caveat that support for the 2016 Mac version of Microsoft Word is promised but unavailable at the time of this writing. MathType is also capable of importing and exporting equations written in LaTeX, although sometimes manual translation can be required for non-standard commands and packages.

Another approach is to use the LaTeX-to-Microsoft Word software by GrindEQ (Acton, ON, Canada), which allows conversion of LaTeX documents into Microsoft Word format, with the equations rendered using either MathType or the Microsoft Word Equation Editor. GrindEQ sells an add-on to Microsoft Word and also offers a per-document translation service online.

The Duxbury Braille Translater (DBT) software suite (Duxbury Systems, Westford, MA) can convert Microsoft Word documents with embedded MathType equations into a native Braille format that renders the equations in Nemeth Code. Our detailed workflow for creating accessible technical documents from source files containing mathematical notation is given below.

**A4.3 Workflow for creating accessible documents that include mathematics and technical notation:**

Convert the file into a Microsoft Word document containing only ASCII characters, with all equations, mathematical symbols, and other non-ASCII characters rendered in MathType (preferably) or if not in Equation Editor. Then:

1. If the file is in LaTeX, use GrindEQ LaTeX-to-Microsoft Word to convert it into a Microsoft Word document with equations rendered in MathType.

2. If the file is already a Microsoft Word document, make sure to replace any non-ASCII symbols in the Microsoft Word document with MathType objects. For example, the character $\pi$ would be substituted with an in-line MathType object that contains only the $\pi$ symbol.

3. For bitmap images (e.g. from a scanner or copier), use Optical Character Recognition (OCR) software to convert the content to text before copying it into Microsoft Word. Adobe Acrobat contains this capability, although more sophisticated OCR applications are also available. Some modern copiers also have the ability to scan to PDFs with OCR text. Whichever OCR method is used, in our experience the OCR conversion typically needs to be double-checked and edited for accuracy. Most OCR programs also do not convert mathematical notation well, so these may need to be copied manually into Microsoft Word equations using MathType.

4. If the file is in another format, such as a PDF whose LaTeX source code is not available, convert the content manually to Microsoft Word by copy-pasting text where possible, and replacing any non-ASCII symbols in the Microsoft Word document with MathType objects.

5. If the Microsoft Word document contains any Equation Editor objects, run the Convert Equations command from the MathType menu in Microsoft Word to convert all of these equations into MathType objects.

6. Now that all mathematical symbols and equations are rendered exclusively in MathType, use Duxbury software to convert the Microsoft Word document into electronic Braille format, with equations expressed in Nemeth Code.

**A4.4 Tactile drawing boards for real-time sketches**

An old and effective method for conveying very simple graphical information is to trace patterns onto the student's hand. Building on this idea, there are currently a range of options for making real-time tactile graphics produced in other contexts where advance printing is not feasible, such as during a lecture session when a diagram is drawn on a chalkboard or whiteboard.

The tactile drawing board we used most often was the DRAFTSMAN (American Printing House for the Blind, Louisville, KY), which can be used with a stylus or a ballpoint pen to produce raised-line drawings in the same time that it would take to draw the diagram for a sighted student. Another relatively low-tech solution is the Sewell E-Z Write N Draw Raised Line Drawing Kit (Maxi-Aids, Farmingdale, NY), which includes a rubber-faced clipboard, aluminum foil sheets, and a stylus. To use it, one can lay the foil over the top of the rubber-faced clipboard and then write or draw on it using the stylus. Having the rubber layer underneath the foil ensures that the foil will not be punctured when drawn on, only indented.

Other tactile drawing boards that allow relatively rapid diagram creation include the InTACT Drawing Bundle (E.A.S.Y. Tactile Graphics, Burlington, VT). The InTACT permits erasing using a small iron that smooths out the surface of plastic paper, and allows others can save drawings electronically as a graphics file. Another option for real-time graphics is the Graph Board, also called the Graphic Aid for Mathematics (American Printing House for the Blind), combined with a Braille or print protractor. The Graph Board is a rubber pad marked with an embossed grid pattern on which plots can be made using rubber bands stretched and held in various shapes using pushpins.

## A4.5 Higher resolution tactile drawings

For more precise tactile plots where the time required to create the drawing is not as critical, there is the TactiPad Drawing Tablet (Thinkable, Huissen, The Netherlands). This device functions similarly to the DRAFTSMAN and Sewell drawing boards, in that it is a plastic framed drawing board that one can place a sheet of thermoform paper on top of and then use a stylus to produce an immediate raised-line drawing on the paper. However, the TactiPad also comes with drawing tools such as a ruler, a compass, and pins, all of which can be used to do more precise tracing when creating tactile diagrams. Additionally, when tracing with a digital pen paired with TactileView Graphics Software, a digital copy of the image is automatically transferred into the software suite and saved on the computer to which it is connected. This image then can be viewed and altered in the software program, as well as printed with an embossing printer (discussed later in this section). Basic images can be created using the TactileView Graphics Software and then transferred into tactile graphics. The TactiPad also allows audio to be attached to the drawing in specific zones so that it can be accessed later when a blind user is interacting with the tactile image.

## A4.6 Tactile pens

Another substitution for the liquid tactile pen is to use a crafting substance, like puffy paint or hot glue that produces raised figures. The end product is durable and can be stored for use after its initial creation. Drawbacks include substantial drying time before the tactile diagram is usable and inherently low-resolution. Newer 3D printing pens that extrude a thin bead of plastic may offer a quick-setting alternative.

### A4.7 Magnet boards

Torres and Mendes (2017) and Supalo et al. (2008) discuss a variety of ways to use magnet boards to represent structures in chemistry and physics. This method could be used to represent free-body diagrams, other vector problems, geometrical optics, electrical circuits and physical phenomena. For all of these mechanical tactile graphics, the resulting outputs can be captured and included in homework or exam solutions with the help of the teaching assistant and a camera.

### A4.8 Braillers (Braille "typewriters")

Braillers act like Braille typewriters that can be used to produce Braille printouts manually. These devices use keys representing the dots for each Braille character, and emboss the corresponding dots and characters onto paper instead of inked letters. Braillers are typically used to print Braille documents manually, but they can also be used to produce tactile plots or line drawings. Since the figures must be produced manually, line-by-line, this process can be time-consuming and difficult. It can also be challenging to produce high-precision or complex figures, because its resolution is typically only 25 lines x 42 cells for an 11.5" x 11" sheet of paper (or 32 cells wide for a standard 8.5" wide sheet of paper). Current Brailler models include the Perkins (Perkins Solutions, Watertown, MA), the Mountbatten (Harpo, Poznan, Poland) and the SMART (Perkins Solutions).

## A4.9 Tactile printers for Braille and graphics

A variety of printers exist that output Braille and other raised renderings for producing tactile printouts. These are made either by embossing Braille- or similarly-sized and spaced dots or by heating special paper to create "swell-form" graphics. Some printers can only print Braille text, while others can also produce tactile graphics (diagrams, plots, etc.) The price and specific features of the available types of Braille printers depend on the purpose that they are intended to serve. The main distinction is between high-volume and low-volume printers. High-volume printers can print about 400 characters per second (CPS), whereas most low-volume printers only print about 40-100 CPS. The majority of the models have a printing resolution of 30-40 characters per line per standard Braille cell and paper sizes. Braille printers also can be connected to a computer or Braille notetaker to allow blind and sighted users to interact in real-time. Ideally, there should be <two embossing printers: one available to everyone involved in course preparation and grading, and a second for the student's independent and personal use.

We primarily used the Emprint SpotDot Braille Embosser (ViewPlus Technologies, Corvallis, OR). When used with the Tiger Software Suite (ViewPlus Technologies), this embosser can output standard Braille as well as Nemeth Braille for mathematics. This combination is compatible with Microsoft Word™ (for creating text and line diagrams) and LoggerPro (for printing tactile graphs). It is also possible to print text that is typed directly in Braille using the installed Braille fonts at 29 point size. Staff typically require about 1 hour of training on how to create and print images using the software and the printer.

Several embossers, including the SpotDot, are capable of producing embossed dots of varying heights, which can be used to emphasize certain aspects of a figure (in much the same way that bold or italics would for a sighted person), or to convey varying topographical information (such as a 2D intensity pattern). The SpotDot and other models can also print in ink on top of the embossing, which allows both sighted and non-sighted people to interact with the same figure. The printed image can also be designed to use a specific color of ink in correlation with a certain texture of embossing so that, in calculus for example, different functions can be printed on the same plot and both the sighted and non-sighted users can understand the content of the image fully. We found that this is particularly useful when our student was working with a tutor, since variations in texture can convey certain information about the image. For example, in calculus, using a lower level of embossed dots to "shade in" the area under a curve can be used to delineate the area that is calculated by taking the integral under the curve.

A swell-form graphic is a raised line drawing that is created by printing or drawing a black-ink image onto special "swell-touch" paper using a standard printer, copier or pen, and then using a special "swell-form" machine to heat the paper. Regions printed with ink then swell so as to produce a raised-line tactile version of the original image. Our institution has used Pictures In a Flash (PIAF) (Humanware) to produce tactile graphics for the student that supplement the material covered in his physics classes. Swell-form printing technology is an efficient way to produce line drawings or other simple tactile graphics. However, the spatial resolution of this technique is lower than high-end embossing printers.

## A4.10 3D representations

It is challenging for a blind student to interpret most 3D diagrams, since the key to perspective drawing lies in using an optical illusion to depict 3D space on a 2D plane. A better approach is to use actual tactile 3D models. For example, pipe cleaners, wooden sticks or building toys such as K'Nex (K'Nex, Hatfield, PA) or Zometools (Zometool Inc., Longmont, CO) can be used to construct a model of the Cartesian coordinate axes and to show the orientation of vectors in 3D. Pipe cleaners and Wikki Stix (a flexible molding rod; Omnicor Inc., Phoenix, AZ) can easily be molded repeatedly into a variety of shapes to represent functional forms, waveforms, and other curves in 2D or 3D. These tools can also be used to show the geometry for projecting vectors onto an axis, creating a cross product of two vectors and other spatial relationships between different vector quantities. Braille labels can distinguish the different vectors; these can either be printed using embossing printers or produced using inexpensive Braille label makers.

To represent functions of two variables, *f(x,y)*, one can print tactile 2D contour plots or create a contour image from multiple layers of art materials (molding compounds, cardboard, etc.) Common household objects can represent complex 3D surfaces: various forms of pasta can stand in many mathematical shapes, springs can represent themselves, a potato chip can be likened to a saddle point, etc. The American Printing House for the Blind also sells a variety of basic mathematical 2D and 3D shapes, including polygons, cones, cylinders and prisms. For modern physics, there are commercial kits available that represent orbitals of the hydrogen atoms, such as Molymod™ (Spiring Enterprises, Billingshurst, West Sussex, UK), which use color and texture to indicate additional features such as phase.

3D printing can be used to produce 3D objects on demand. Online file-sharing websites such as Thingiverse (www.thingiverse.com) are making descriptor files available for download to facilitate this process.

## A4.11 Other tactile and haptic technologies

A variety of emerging technologies hold great promise for the future. Tablet computers can allow users to examine graphical information via touch so as to evoke audible information precoded into the graphic. However, we were not able to find suitable physics-based content for these tools. Haptic (touch-based) devices such as data gloves that provide force or vibrational feedback promise to provide a way to simulate the feel of complex 3D shapes or to query datasets in two or three dimensions using hand gestures.

## A.5.1  Mathematica accessibility resources

Kyle Keane, PhD (Education and Accessibility Consultant, Wolfram Research Inc.) recommends these resources for learning how to use wolframscript interface to interact with Mathematica:

- http://reference.wolfram.com/language/XML/tutorial/MathML.html

- http://reference.wolfram.com/language/howto/ImportAndExport.html

- https://www.wolfram.com/wolframscript/

- http://reference.wolfram.com/language/ref/program/wolframscript.html

- http://reference.wolfram.com/language/tutorial/WolframLanguageScripts.html

- http://reference.wolfram.com/language/ref/format/SVG.html

Users interested in more information about accessible use of Mathematica can post questions and get support by emailing [support@wolfram.com](mailto:support@wolfram.com) or by reaching out on the community forum at:

[http://community.wolfram.com/content?curTag=accessibility](http://community.wolfram.com/content?curTag=accessibility)